# A Bibliography of Combinators

Stephen Wolfram*

*A categorized bibliography of combinators is given, providing what is believed to be a largely complete coverage of publications from the origination of combinators in 1920 to the present day.*

## Foundational Documents

## Books

*Email: s.wolfram@wolfram.com

## Surveys & Summaries

# Combinators as Symbolic Expressions

## Specific Combinators & Behavior

## Conversions & Notations

## Combinator Reduction

## Random Combinators

# Combinators as Mathematical Constructs

## Combinatory Logic

## Models of Combinatory Logic

## Relations to Lambda Calculus

## Relations to Type Theory

## Relations to Recursive Functions

## Relations to Other Mathematical Structures

# Combinator Computation

## Combinator Evaluation

## Compilation to Combinators

## Combinators in Functional Programming

## Metaprogramming with Combinators

## Specific Programming Tasks

# Extensions & Applications

## Extensions of Combinators

## Combinatory Grammars & Linguistics

## Confusing Issues

*The term "combinatory analysis" has nothing to do with "combinators"; it's an earlier name for "combinatorics", used for example in:*

P. A. MacMahon (1915), *Combinatory Analysis*, The University Press (Cambridge).

*"Combinatory" is also not used in the sense of combinators in:*

E. Post (1936), "Finite Combinatory Processes—Formulation 1", *The Journal of Symbolic Logic* 1, 103–105. doi: 10.2307/2269031.

*"Combinator" is sometimes used as a fairly general term for a function or operation that combines computational operations, as in:*

R. Milner (1982), "Four Combinators for Concurrency", in *Proceedings of the First ACM SIGACT-SIGOPS Symposium on Principles of Distributed Computing*, Symposium on Principles of Distributed Computing, Association for Computing Machinery, 104–110. doi: 10.1145/800220.806687.

L. Cardelli and R. Davies (1999), "Service Combinators for Web Computing", in *Transactions on Software Engineering*, IEEE, 309–316. doi: 10.1109/32.798321.

*"The Combinator" is a recent combinatorial tool for idea generation, that seems to have no relation to combinators:*

J. Han, et al. (2018), *The Combinator—A Computer-Based Tool for Creative Idea Generation Based on a Simulation Approach*, Cambridge University Press. doi: 10.1017/dsj.2018.7.

*Y Combinator is a startup accelerator founded by P. Graham et al. in 2005; its name is derived from the fixed-point combinator, but otherwise it is unrelated:*

Y Combinator (accessed March 22, 2021). www.ycombinator.com.

## Thanks

Thanks to Henk Barendregt, Ariela Böhm, Emanuele Böhm, Michele Böhm, Martin Bunder, Mariangiola Dezani-Ciancaglini, Silvia Ghilezan, Roger Hindley, Oleg Kiselyov, Jan Willem Klop, Gerd Mitschke, Alberto Pettorossi and Adrian Rezus for suggestions and material for this bibliography, and to Paige Bremner, Amy Simpson and the University of Illinois library for extensive work in tracking down documents.

## References

*Links to references are included within the body of this document.*

*Cite as:* S. Wolfram (2021), "A Bibliography of Combinators". wolframcloud.com/obj /sw-writings/Combinators/bibliography.pdf.